# Neurodevelopmental Age Estimation of Infants Using a 3D-Convolutional Neural Network Model based on Fusion MRI Sequences

Mahdieh Shabanian[1,2,*], Adeel Siddiqui[3], Hao Chen[4], John P. DeVincenzo[5,2]

[1]Department of Biomedical Engineering, University of Tennessee Health Science Center, Memphis, TN, USA
[2]Children's Foundation Research Institute at LeBonheur Children's Hospital, Memphis, Tennessee, USA.
[3]Department of Radiology, LeBonheur Children's Hospital, University of Tennessee Health Science Center, Memphis, TN, USA
[4]Department of Pharmacology, Addiction Science, and Toxicology, University of Tennessee Health Science Center, Memphis, TN, USA
[5]Departments of Pediatrics and Microbiology, Immunology, and Biochemistry, University of Tennessee School of Medicine, Memphis, TN, USA

## Abstract

**The ability to determine if the brain is developing normally is a key component of pediatric neuroradiology and neurology. Brain magnetic resonance imaging (MRI) of infants demonstrates a specific pattern of development beyond simply myelination. While radiologists have used myelination patterns, brain morphology and size characteristics in determining if brain maturity matches the chronological age of the patient, this requires years of experience with pediatric neuroradiology. Due to the lack of standardized criteria, estimation of brain maturity before age three remains fraught with interobserver and intraobserver variability. An objective measure of brain developmental age estimation (BDAE) could be a useful tool in helping physicians identify developmental delay as well as other neurological diseases. We investigated a three-dimensional convolutional neural network (3D CNN) to rapidly classify brain developmental age using common MRI sequences. MRI datasets from normal newborns were obtained from the National Institute of Mental Health Data Archive from birth to 3 years. We developed a BDAE method using T1-weighted, as well as a fusion of T1-weighted, T2-weighted, and proton density (PD) sequences from 112 individual subjects using 3D CNN. We achieved a precision of 94.8% and a recall of 93.5% in utilizing multiple MRI sequences in determining BDAE.**

## Introduction

Neurodevelopmental disorders (NDDs) are a diverse group of conditions characterized by delayed milestones involving cognition, communication, behavior, and motor skills. Sometimes, the underlying neurodevelopmental abnormality is apparent. However, magnetic resonance imaging (MRI) scans of these patients often reveal normal imaging and neurological findings.

Infancy and early childhood are characterized by rapid cognitive development, especially in the first three years of life. This cognitive development is mirrored by changing brain structure, function, and connectivity. Many pediatric diseases impair this development. Therefore, brain developmental age estimation (BDAE) is crucial to determine if a child's brain is developing normally. Neuroimaging of infants offers qualitative information such as myelination patterns and brain morphology. MRI also offers quantitative information such as head circumference, brain volume, and water content. However, this quantitative information is difficult for radiologists to extrapolate. Neuroimaging-derived brain age is critical for determining the impact of pediatric brain diseases. Healthy clinical sample brain-MRI data can be used to train a deep-learning program for healthy infant brains. This standard could permit comparisons of normal brain development with many infant diseases, including prematurity [1], hypoxic ischemic encephalopathy (HIE) of the newborn [2], congenital cytomegalovirus (cCMV) infection [3], bacterial meningitis, herpes simplex virus encephalitis (HSVE) [4], pediatric epilepsy, cerebral palsy (CP), and genetic disorders.

Newborns with brain disorders are at increased risk of severe long-term motor deficit and cognitive delay. Detecting these conditions early could help mitigate morbidity. Based on analyses of magnetic resonance (MR) images, the brain shows specific morphological changes throughout the first three years of life; variations in these morphological changes could indicate abnormal development. Volumetric changes in brain tissues, such as white matter (WM), gray matter (GM), and cerebrospinal fluid (CSF), are well documented but difficult to apply in clinical practice [5, 6, 7, 8, 9]. For example, calvarial growth outpaces brain growth in the first three years of life, leading to enlarged subarachnoid CSF spaces. Later on, in life, the brain starts matching calvarial growth and the caliber of these CSF spaces decreases. This interplay between brain growth and calvarial growth is assessed on a qualitative basis but has yet to be studied in a quantitative basis using artificial intelligence algorithms.

Normal brain development may be disrupted in premature infants, infants with genetic diseases, and infants with infections. Early diagnosis of neurodevelopmental delay would help physicians manage patients and decide which treatment algorithms to pursue. It would help doctors counsel parents about what to expect as their child gets older. BDAE may become an important adjunct screening in patients with mild cognitive impairment or a tool to monitor neurodevelopment in patients with diseases such as tuberous sclerosis and cystic fibrosis.

Adult-optimized imaging parameters are difficult to apply to infants. In fact, both acquisition and computational approaches in neonatal imaging have yet to achieve an acceptable level of consensus. The acquisition of neonatal MRI poses an additional challenge because they have a lower signal to noise ratio (SNR) due to their smaller brains and neonates are more susceptible to movement in the loud MRI environment. Moreover, the large morphological variability observed in clinically abnormal populations, especially pathology leading to ventriculomegaly and white matter lesions, requires more flexible algorithms.

The neonatal brain shows a much higher water content at birth, with a subsequent rapid decrease in water content. In the first year of life, myelination occurs very rapidly with GM migrating from its origin in the periventricular region to the cortex. These properties change rapidly throughout the first few years of development and lead to many computational challenges not seen in adult models. As a result, neonatal development as observed by MRI can be roughly divided into four distinct temporal stages [10].

The first stage is the infantile/neonate pattern (< 4 months) and is a reversal of the grey-white tissue contrast observed in the adult pattern (T1w and T2w are effectively switched). The majority of the WM is T2-weighted hyperintense, making it difficult to differentiate from cerebral edema in the neonatal period. The second stage (between 5-9 months of age) is characterized by further progression of myelination anteriorly, superiorly and laterally [11]. The splenium of the corpus callosum becomes myelinated during this time. The third stage (around 12 months) is reached once the rate of myelination drops off, with tissue contrast more closely resembling a fully developed brain on T1-weighted imaging. The fourth stage (1-2 years) is essentially the adult myelination pattern on T1-weighted imaging, however T2-weighted sequences will still have areas of bright signal that are not found in an adult.

With no biomarkers to diagnose NDD or quantitative assessment of neurological pathology in infants, these NDDs are identified through physical exams and imaging tests. Advances in MRI have enabled the noninvasive visualization of the infant brain through high-resolution images. However, the processing and analysis of MR infant brain images are typically more challenging than those of adult brain images.
Preterm infants are at high long-term risk of severe neurodevelopmental delay and cognitive deficit. Several systematic reviews of long-term neurodevelopmental outcomes after intrauterine and neonatal insults reported the preterm neonates present with multiple insults including neonatal sepsis, meningitis, hypoxic ischemic encephalopathy, jaundice, tetanus and congenital infections such as cytomegalovirus, herpes, rubella and toxoplasmosis [12, 13, 14, 15]. Therefore, accelerating the identification of impairment is needed in infants to prevent long term abnormalities and improve progression.

Advancements and improvements in medicine and healthcare in the past few decades have ushered in a data-driven era where abundant data is collected and stored. With this change, there is a great need for an analytical and technological upgrade of existing systems and processes. Medical data, such as images, demographic information, medication history, or patient visits, are becoming universally digitized. Other than rapid access across institutions, this knowledge is largely underutilized. Automated BDAE algorithms for the early diagnosis of these infants would be helpful. Leveraging artificial intelligence algorithms makes it possible to evaluate early signs of delayed age objectively and quantitatively for each patient. Machine learning is the study of models that computer systems use to self-learn instructions based on the weight of parameters without explicit instructions. In parallel with biomedical advancements in the past decade, increased refinement of algorithms and machine learning tools has occurred. Deep learning (DL), one of the more promising algorithms, is an artificial neural network that designs models computationally. DL is comprised of many processing layers and able to learn data representations with many levels of abstraction [11, 16, 17, 18]. DL algorithms show exciting potential in learning patterns and extracting attributes from complex datasets in health informatics and biomedicine. BDAE classical approaches using manual extraction of features such as machine learning algorithms may not fully represent the MR images contents [19, 20, 21]. We hypothesized DL approaches could be used to classify normal age specific MRI brain development accurately based on MRI sequences. This normative classification could then be applied as a BDAE score to identify patients with developmental delay and serve as an objective finding that could help eliminate interobserver and intraobserver variability. This score could also be used to risk-stratify different patient populations.

# Deep learning

Traditionally, machine learning models are trained to perform useful tasks based on features manually extracted from raw data or are engineered using other classical machine learning models. In DL, computers learn high-level features automatically, directly from raw data, bypassing an often difficult feature engineering step. Automatic feature detectors represent a significant difference between DL algorithms and more classical machine learning. Yann et al. published a review paper titled "Deep Learning" in Nature to understand the fundamental strengths of deep learning algorithms [22]. Litjens et al. also provided a general overview of deep learning algorithms in radiology [23].

DL algorithms, specifically convolutional neural networks (CNNs), have become widely used for analyzing medical images. In medical imaging, the interest in DL is mostly driven by CNNs, which can uncover useful unseen features of images. With a deep neural network, many image features are typically preserved in image analysis. Some potent preferences are embedded in CNNs [24, 25], explaining how CNNs provide powerful image analysis while lessening pre- and post-processing tasks to extract higher level features to reduce an image to its key features, thus enabling easier classification [26].

Various types of CNNs were developed recently, and all have the potential to contribute to the accuracy and speed of image classification or automatic segmentation. CNNs can learn high-level hierarchies of features automatically by backpropagating errors through multiple blocks and layers, such as convolutional layers, max-pooling layers, and fully connected layers used for image classification tasks [24].

## A. Structure of Convolutional Neural Networks

CNNs are able to learn spatial hierarchies of features automatically and adaptively using multiple building blocks, such as convolutional layers, pooling layers, and fully connected layers. In fact, CNN is trained through backpropagation and gradient descent to generate an error signal that measures the difference between the estimation of the network and the target value. CNN uses this error signal to change the weights or parameters for more accurate classification or segmentation [27]. CNNs typically have several convolution layers, pooling layers, and fully connected (FC) layers at the end, which compute the final outputs (Figure 1).

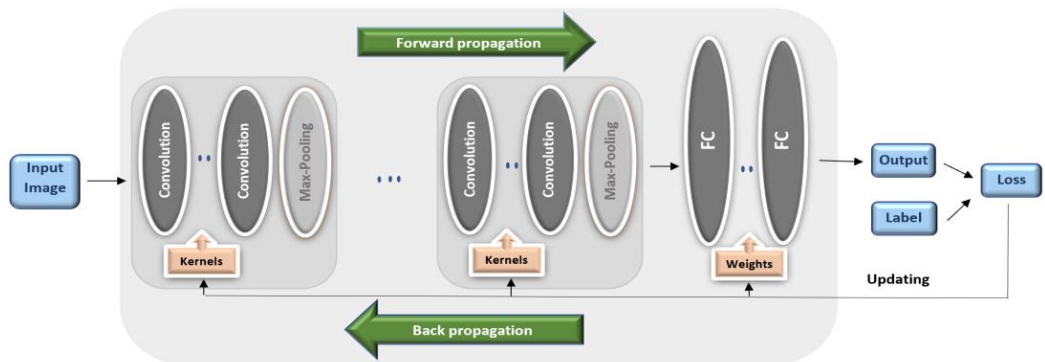

Figure 1. An overview of a convolutional neural network (CNN) architecture and the training process. The performance of the model is calculated under weights and kernels with a loss function through forward propagation on a training dataset. All learnable parameters, such as weights, kernels are updated based on the loss value through backpropagation with gradient descent optimization algorithm [25]. FC: fully connected layer.

## B. Hardware and Software

Training CNNs usually require a graphic processing unit (GPU) with memory matched to the number of parameters in the network. GPUs led to the remarkable rise of DL. Technically, GPUs increase speed by 10 to 50 times of CPU-based work for training deep neural networks. The AlexNet success developed by Krizhevsky et al. is due to its architecture [28], the power of calculation is based on the GPU.

Several open-source DL libraries provide efficient GPU implementations of deep neural networks, such as convolutions [22]. In 2015, Google Research developed TensorFlow [29], which provides C++ and Python interfaces and is used by Google AI. Google's engineer developed Keras, a popular framework that provided the Python interface in 2017.

## C. Deep Learning Algorithms in Medical Image Analysis

Many image diagnosis tasks require an initial search to identify abnormalities. Measurement of lesions is often done, which can have significant variability between radiologists. If there are prior images, then any changes over time is closely examined. Automated medical image analysis tools based on DL algorithms could be helpful tools in adding objective, algorithmic elements to image interpretation. This approach has opened new doors in medical image analysis. DL applications in medical image analysis cover a broad spectrum of modalities, including X-Ray, CT and MRI. There have been comprehensive surveys of deep learning in medical imaging [22, 25, 30].

## D. 3D CNN Model

The development of 3D CNNs remains at an early stage because of their complexity, especially in medical image interpretation. Around 2018-2019, exponential growth was seen in using 3D deep learning models in medical imaging [31]. This review paper also described the development of 3D CNN from machine learning roots in detail. Convolutions of 2D CNNs are used to derive features only from spatial dimensions in 2D feature maps. 2D CNNs take a 2D matrix as an input, such as image slices. Therefore, information from adjacent slices is unavailable. With this approach, information is more likely to be lost from interest regions of the brain in specific diseases because the brain is a 3D structure. Figure 2a shows the 2D filter (kernel) in 2D CNN used to extract spatial-spectral features.

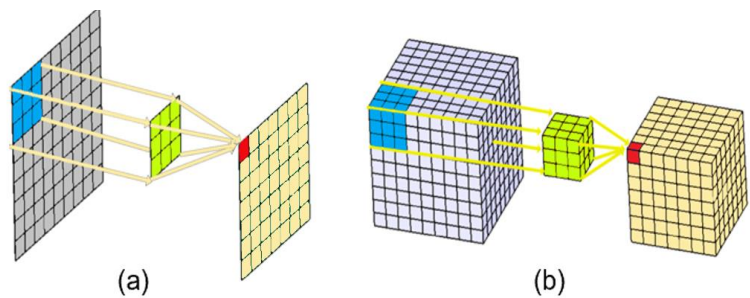

Figure 2. Illustration of convolution: (a) illustration of a 2D filter to extract spatial features; (b) illustration of a 3D filter to extract spatial-spectral features.

The 3D CNN architectures using volumetric convolutions are very useful DL methods for analyzing volumetric imaging data. The 3D CNN mathematical model is very similar to 2D CNN by adding one extra dimension. Volumetric image analysis is a time-consuming process that currently requires expert knowledge, especially for medical images such as those MRI images using the Neuroimaging Informatics Technology Initiative (NIfTI) formats [32, 33, 34]. For the first time, Cole et al. showed that 3D-CNN could accurately estimate developmental brain age from MRI data using healthy adult samples [34]. Additionally, authors demonstrated 3D CNNs are more effective and less likely to miss regions of interest in medical images.

3D convolution applies a 3D filter (kernel) to the dataset, and this filter moves in 3-directions (x, y, z) to calculate low-level feature representations (Figure 2b). 3D CNNs help prevent loss of the region of interest that occurs in 2D medical images. CNNs apply a filter to the input to create a feature map to summarize detected features in the input. Filters, such as line detectors, edges, and curved lines can be created, but the CNNs innovation is to learn the filters during training in the context of a multi classification problem. Statistical tests indicated that the 3D CNN model was able to classify 2D MRIs more accurate than the 2D CNN model [26]. 3D CNNs can use entire image volumes rather than individual slices such as 2D CNNs, allowing for more information (e.g., relationships between consecutive slices). Implementation of 3D models can also be more difficult on commonly available GPU cards due to their large memory consumption. A 3D model has better performance when sufficient training data and powerful computational hardware are available. Several articles demonstrated the better performance of 3D CNN on MRI brain scans [26, 35, 36, 37].

Using 3D CNN models, we can now observe subtle micro- and macrostructural changes, as well as aberrant connectivity variances that may go undetected at an age where early intervention could have impacted patient care. In addition, these findings may be undetectable by the human eye until much later in the child's development, preventing implementation of an early treatment plan.

## Methods

### A. Dataset

We obtained 570 normal MRI scans (T1w, T2w, and PDw sequences) of 112 infants from the NIMH Data Archive (NDA). In this dataset, we had 190 T1-weighted sequences, 360 scans for both T2-weighted and PD-weighted sequences. The NIMH pediatric MRI images came from infants and young children ranging in age from 8 days to 3 years. All infants were screened for the presence of factors that might adversely affect brain development (e.g., preterm birth, learning disabilities, physical growth delay, perinatal complications, and psychiatric disorders of first-order family members).

The infants were scanned with 1.5T MRI while awake or during natural sleep without sedation. MRI acquisition generally lasted 30–45 minutes on a 1.5T scanner with a 2D sequence that minimized scan duration for the birth to 3 years age-groups. The axial scans consisted of a 2D T1-weighted spin echo and a T2-weighted 2D Fast Turbo spin echo sequence. The T1-weighted sequences utilized TR 500, TE 12, 90-degree flip angle, and 3mm slice thickness. The T2-weighted and PD-weighted scans utilized a TR 3500,

TE 15-17 (115-119), and a 3 mm slice thickness. The T1-weighted and T2-weighted scans were nominally 1×1×3 resolution (1×1×3 or .97×.97×3) with a matrix of 256 x 192 mm. The majority of the scans were provided by Siemens Medical Systems (Sonata, Magnetom) scanner, while another site used a GE (Signa Excite) scanner to provide less than half of the scans [38].

We previously developed a neural network-based method for classifying three age groups [26]. This dataset allows us to refine our method to classify subjects into much finer age groups that may be useful for clinical applications. This cohort age included: 2 weeks (8 to 35 days, n=21), 3 months (each ± 2-weeks, n=34), 9 months (each ± 2-weeks, n=31), 12 months (each ± 2-weeks, n=31), 24 months (each ± 2-weeks, n=33), and 36 months (each ± 6-weeks, n=55). In this study, we analyzed and classified these six age cohorts with respect to their brain MRI. Each subject had T1-weighted, T2-weighted, and PD-weighted sequences except for some subjects who had only two of three MRI modalities in the dataset. Some of the subjects had several visits at different ages.

## B. Measuring Metrics

The statistical measures to evaluate the performance of multi classification methods are described below. True positive (TP), false positive (FP), true negative (TN), and false negative (FN) values are used for calculating performance measures. TP denotes the number of instances belonging to a class that are recognized accurately as the members of that class. FP denotes the number of instances incorrectly recognized as the members of a class that belong to other classes. TN is the number of instances correctly recognized as not belonging to a class, and FN is the number of instances incorrectly recognized as not belonging to a class. Equations 1-4 calculate the accuracy, recall, precision, and F1 score of the BDAE system.

$$Accuracy = \frac{(TP+TN)}{(TP+TN+FP+FN)} \tag{1}$$

$$Recall = \frac{TP}{(TP+FN)} \tag{2}$$

$$Precision = \frac{TP}{(TP+FP)} \tag{3}$$

$$F1 - score = \frac{(2*Precision*Recall)}{(Precision+Recall)} \tag{4}$$

The correct percentages and BDAE classification errors are shown in the normalized confusion matrix.

## C. Experiments and Results with T1-Weighted Images

Several studies used CNN to estimate brain age using T1-weighted MRI volumes [32,34,35]. In this study, 3D CNN was used for brain developmental age estimation in six major age cohorts in infants from birth to 3 years. This study was carried out in two stages. In the first stage, we used 190 T1-weighted sequences using proposed 3D CNN for age classification. In the second stage, we used a fusion of T1-weighted, T2-weighted, and PD-weighted MRI sequence images to estimate and classify these six age cohorts.

We used Keras (version 2.2.4) as the software framework. A computer with an NVIDIA TITAN RTX GPU performed all calculations. All NIfTI MRIs were rescaled to a standard size (80, 80, 80 voxels) with gray scale for 3D models to reduce computational complexity. The training set and validation set were randomly split as 80% and 20% of the dataset, respectively.

**3D CNN Model with T1-Weighted MRI Sequences**

Figure 3 shows the 3D-CNN architecture used in the proposed brain developmental age estimation methodology, the same as in our previous research [26] using T1-weighted scans as an input with low resolution (80, 80, 80 voxels). This proposed architecture has 4 convolutional blocks consisting of a 3D convolutional layer (kernel size: 3×3×3). The number of feature channels for each block is 32, 64, 128, and 256. The last three layers are fully connected layers for combining the feature vectors. The output is an estimated age of six age cohorts. Due to its superior grey-white matter differentiation, T1-weighted MRI has a crucial role in estimating age in an infant. In this model, cross entropy, RMSprop and rectified linear (ReLU) were used for the loss function, optimizer, and activation functions, respectively. Commonly, cross entropy is used for loss function in multiclass classification. The learning rate is another important hyperparameter, which we chose 0.0001. We tried various optimizers, kernel sizes, and learning rates and selected the best hyperparameters for our 3D CNN. Table 1 shows in detail all layers in the proposed 3D CNN model with batch normalization and dropout.

| Layer (type) | Output Shape | Param # |
|---|---|---|
| conv3d_1 (Conv3D) | (None, 80, 80, 80, 32) | 896 |
| batch_normalization_1 | (None, 80, 80, 80, 32) | 128 |
| max_pooling3d_1 | (None, 80, 40, 40, 32) | 0 |
| conv3d_2 (Conv3D) | (None, 78, 38, 38, 64) | 55360 |
| conv3d_3 (Conv3D) | (None, 76, 36, 36, 64) | 110656 |
| conv3d_4 (Conv3D) | (None, 74, 34, 34, 64) | 110656 |
| conv3d_5 (Conv3D) | (None, 72, 32, 32, 64) | 110656 |
| batch_normalization_2 | (None, 72, 32, 32, 64) | 256 |
| max_pooling3d_2 | (None, 72, 16, 16, 64) | 0 |
| dropout_1 (Dropout) | (None, 72, 16, 16, 64) | 0 |
| conv3d_6 (Conv3D) | (None, 70, 14, 14, 128) | 221312 |
| conv3d_7 (Conv3D) | (None, 68, 12, 12, 128) | 442496 |
| batch_normalization_3 | (None, 68, 12, 12, 128) | 512 |
| max_pooling3d_3 | (None, 68, 6, 6, 128) | 0 |
| dropout_2 (Dropout) | (None, 68, 6, 6, 128) | 0 |
| conv3d_8 (Conv3D) | (None, 67, 5, 5, 256) | 262400 |
| conv3d_9 (Conv3D) | (None, 66, 4, 4, 256) | 524544 |
| conv3d_10 (Conv3D) | (None, 64, 2, 2, 256) | 1769728 |
| batch_normalization_4 | (None, 64, 2, 2, 256) | 1024 |
| max_pooling3d_4 | (None, 64, 1, 1, 256) | 0 |
| dropout_3 (Dropout) | (None, 64, 1, 1, 256) | 0 |
| flatten_1 (Flatten) | (None, 16384) | 0 |
| dense_1 (Dense) | (None, 1024) | 16778240 |
| dropout_4 (Dropout) | (None, 1024) | 0 |
| dense_2 (Dense) | (None, 512) | 524800 |
| dropout_5 (Dropout) | (None, 512) | 0 |
| dense_3 (Dense) | (None, 6) | 3078 |

```
Total params: 20,916,742
Trainable params: 20,915,782
Non-trainable params: 960
```

Table 1. 3D CNN model.

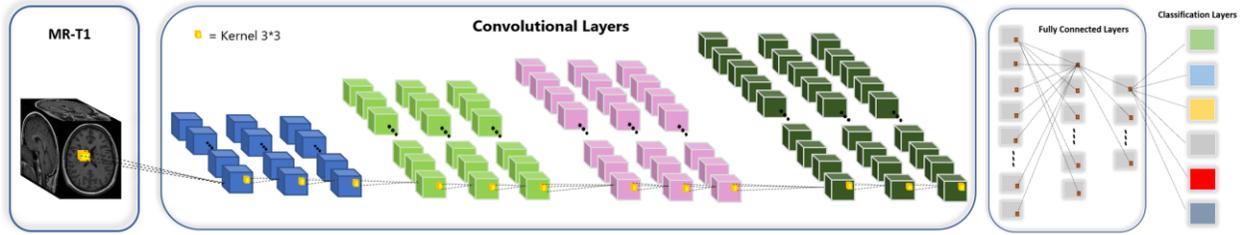

Figure 3. The 3D convolutional neural network architecture. 3D boxes represent input and feature maps. The proposed neurodevelopmental age estimation architecture contains 4 blocks of 3D convolutional operation with a 3D batch normalization, ReLU and, a max-pooling operation after each block of 3D convolutional layers. Three layers of a fully connected layer at the end, that generate the regression model to output predicted brain age in six age cohorts.

We used 152 T1w scans for training and 38 T1w scans for validation (test). In total, we used 36,480 slices of axial, sagittal, and coronal for training and 9,120 slices for validation.

Overfitting in small datasets remains a concern. When training large neural networks from limited training data, the over-fitting problem needs to be considered. Unfortunately, we usually lack access to big data in medical imaging, such as infants' normal MRI. Overfitting occurs when a network learns a function with very high variance, such as perfectly modeling the training data. The best way to recognize overfitting is to monitor the loss on the validation and training sets during the training iteration. If the model performs better on the training set than on the validation set, then the model has been overfitting to the training data.

We achieved 99% accuracy in the training dataset and 50% accuracy in validation test (Supplementary Table 1). Supplementary Figures 1 and 2 show the result while overfitting. Several methods have been proposed to minimize overfitting. Obtaining more training data is the best solution for reducing overfitting by augmentation methods. Data augmentation is another technique for reducing overfitting using deep learning algorithms. We increased the number of T1-weighted scans by flipping and shifting to the right and left, such as common augmentation techniques in medical imaging. Then, we obtained 608 T1w scans (145,920 slices) for training and 152 T1w scans (36,480 slices) for validation. Additionally, dropout and batch normalization methods were used in this section to improve the performance of our 3D model.

Dropout is a regularization technique that zeros out the activation values of randomly chosen neurons during training [39]. The key objective is to randomly drop units (along with their connections) from the neural network during training. This approach reduces overfitting significantly and provides major improvements over other regularization methods. For example, dropout improves the performance of neural networks on supervised learning tasks in vision [39].

The batch normalization technique is for training very deep neural networks that standardize the inputs to a layer for each mini batch. This approach stabilizes the learning process and dramatically reduces the number of training epochs required to train deep networks [40,41]. The batch normalization technique places all the activation values on the same scale.

As seen in Figure 4, we addressed the overfitting issue and achieved 82% accuracy in test using T1-weighted to classify six age cohorts in infants. Table 2 provides the recall and precision in each age cohort and the number of each T1-weighted scans used as test randomly. Figure 5 shows a normalized confusion matrix of our proposed 3D CNN in each class after augmentation of the dataset in stage one with T1-weighted scans. The confusion table shows the correct percentages in each class and BDAE system errors.

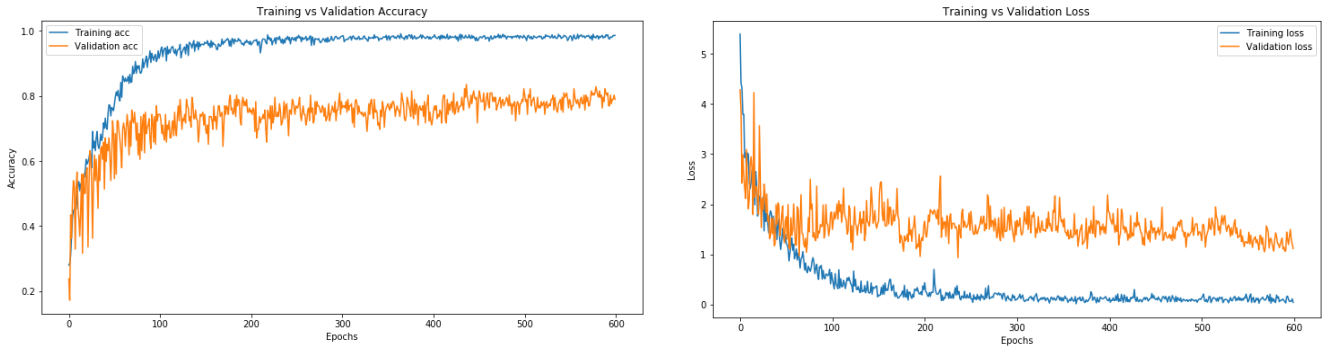

Figure 4. Accuracy and loss in training and validation after 600 epochs using 760 T1w sequences.

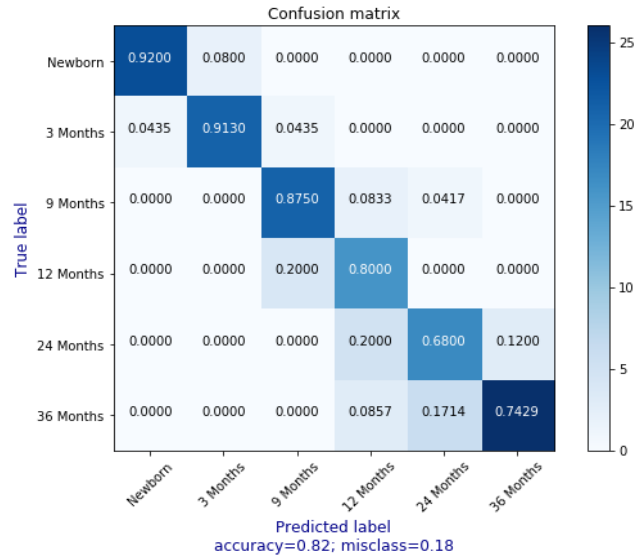

Figure 5. Relationship between neurodevelopmental age and predicted brain age using 760 T1w sequences.

| Class | Statistical Result of 3D CNN | | | Support |
|---|---|---|---|---|
| | *Precision* | *Recall* | *F1-score* | *N Scans* |
| newborn | 0.96 | 0.92 | 0.94 | 25 |
| 3 Months | 0.91 | 0.91 | 0.91 | 23 |
| 9 Months | 0.81 | 0.88 | 0.84 | 24 |
| 12 Months | 0.62 | 0.80 | 0.70 | 20 |
| 24 Months | 0.71 | 0.68 | 0.69 | 25 |
| 36 Months | 0.90 | 0.74 | 0.81 | 35 |

Table 2. 3D CNN statistical results after 600 epochs using 760 T1w sequences.

## D. 3D CNN Model with Fusion of MRI Sequences

Figure 6 shows the same 3D-CNN architecture using the early fusion strategy 570 MR-T1w, T2w and PDw. Before the first convolutional layer, all MR-T1w, T2w and PDw scans are combined before the first convolutional layer are combined as inputs in same directory, and feature maps from every convolutional layer are connected to the first fully connected layer. As above, cross entropy and RMSprop were used for the loss function and optimizer, respectively. The learning rate chose 0.0001. We trained the proposed 3D CNN model with the fusion of T1-weighted, T2-weighted, and PD-weighted in each age cohort. We trained the proposed 3D CNN model by 570 fusion MRI sequences. We used 456 scans (109,440 slices) for training and 114 scans (27,360 slices) for validation. We did not use the augmentation technique in the fusion MRI dataset.

Figure 7 compares the accuracy and loss function after 600 epochs in training and validation. Figure 8 shows a normalized confusion matrix of our proposed 3D CNN in each class. Table 3 shows the recall and precision in each class. As we mentioned above, dropout is a good way to reduce the error on the validation set by averaging the predictions produced by many different networks. Supplementary figures 4-7 and tables 1-3 show step by step performance improvement via dropout and batch normalization. Finally, we achieved 99% accuracy on the training dataset and 94% on the test dataset. We achieved better performance in validation by fusion of MRI sequences. Despite the low resolution of these datasets, our 3D CNN model achieved high accuracy in estimating brain age.

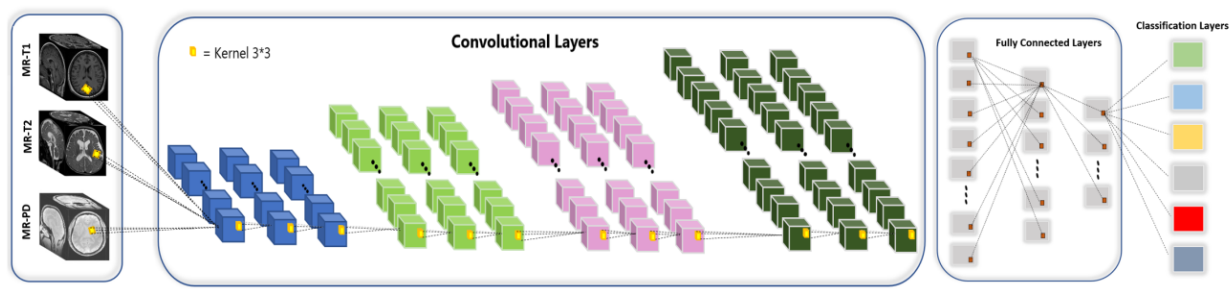

Figure 6. Proposed 3D convolutional neural network architecture using the early fusion strategy. MR-T1w, T2w and PDw scans are combined before the first convolutional layer and feature maps from every convolutional layer are connected to the first fully connected layer.

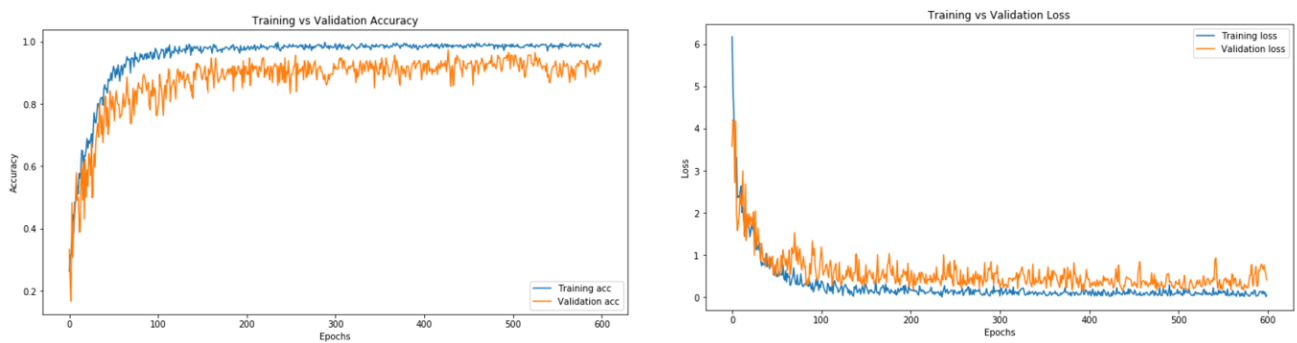

Figure 7. Accuracy and loss in training and validation after 600 epochs using 570 fusion MRI sequences.

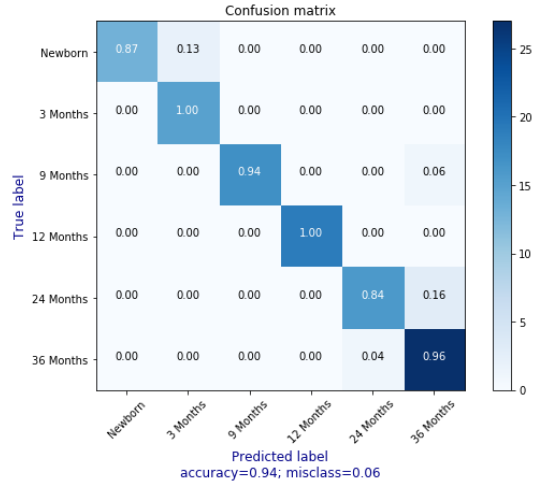

Figure 8. Relationship between neurodevelopmental age and predicted brain age in a 3D classification model after 600 epochs using 570 fusion MRI sequences.

| Class | Statistical Result of 3D CNN | | | Support |
|---|---|---|---|---|
| | *Precision* | *Recall* | *F1-score* | *N Scans* |
| newborn | 1.00 | 0.87 | 0.93 | 15 |
| 3 Months | 0.88 | 1.00 | 0.94 | 15 |
| 9 Months | 1.00 | 0.94 | 0.97 | 18 |
| 12 Months | 1.00 | 1.00 | 1.00 | 19 |
| 24 Months | 0.94 | 0.84 | 0.89 | 19 |
| 36 Months | 0.87 | 0.96 | 0.92 | 28 |

Table 3. 3D CNN statistical results after 600 epochs using 570 fusion MRI sequences.

## Summary of the statistical results

Table 4 provides a summary of the statistical results of the proposed 3D CNN model in two stages. We achieved higher accuracy in fusion MRI sequences.

| Model | Statistical Results | | | |
|---|---|---|---|---|
| | *Precision* | *Recall* | *Accuracy* | *Number of scans* |
| 3D CNN+ Augmentation+ Batch Normalization +Dropout | 81.83% | 82.16% | 82% | 760 (T1w sequences) |
| 3D CNN+ Batch Normalization+ Dropout | 94.83% | 93.5% | 94% | 570 (T1w, T2w, and PDw sequences) |

Table 4. Summary of statistical results with 760 T1w sequences and 570 fusion MRI sequences.

## Limitations and Conclusion

Many pediatric disease processes impair brain development during the first three years of life. Brain MRI holds great promise for studying infant brain development. However, access to this information is limited by subjective human interpretation and requires substantial resources for analysis. There is a lack of advanced digital tools for brain MRI analysis that leverages deep learning. Neuroradiologists and their consulting physicians may be missing a lot of hidden pathology by not fully taking advantage of this resource. Post-processing represents another limitation in clinical applications due to time involved and difficulty in matching results across different institutions. By contrast, clinical decisions are needed in minutes or less and require highly specialized expertise developed over years of practice. Early NDD prediction is a major goal in precision medicine and could be achieved through the application of automated intelligent algorithms. The growing popularity of artificial neural networks and deep learning in medical imaging demonstrates how these tools can accelerate classification in areas such as neuroimaging.

Despite this need, CNN and deep learning has a long way to go. The primary limitation of this paper was not enough cases and sequences were present in the NIH dataset to provide for robust training and validation. Another limitation was the inability of the kernel to preserve matrix size, and the need to rescale NIfTI images. A third limitation was that since this was a publicly available research dataset, the exact date of birth could not be obtained and more precise characterization of BDAE was not possible. Despite, these findings, the high accuracy suggests that BDAE using 3D CNN may have a role in patient care. Repeating and improving these findings using a larger data set of patients, kernels, sequences, and modalities is currently under way. Comparing those results to the patient's exact date of birth will give us a greater understanding as to how precise BDAE can become.

To our knowledge, this study is the first to demonstrate that 3D CNNs can be used to accurately estimate neurodevelopmental age in infants based on the fusion of brain MRIs. Our approach provided accurate results from T1-weighted, T2-weighted, and PD-weighted MRI brain sequences of healthy infants from the NIHPD dataset. In a common MRI dataset, our 3D CNNs obtained 94% accuracy in estimating neurodevelopmental age with minimal processing. Based on these results, 3D CNNs could be used to determine the trajectory of normal brain development and neurodevelopmental age within the first three years of life. This approach could also prove useful in identifying otherwise undetectable abnormalities in brain development.

## Acknowledgments

We would like to express our special thanks to the NIMH Data Archive for access to the NIH pediatric MRI dataset. We would also like to thank Dr. Courtney-Bricker and Ms. Hines for copy editing and proofreading the manuscript.

## Competing interests

The authors declare no competing interests.

## Supplementary Figures and Tables

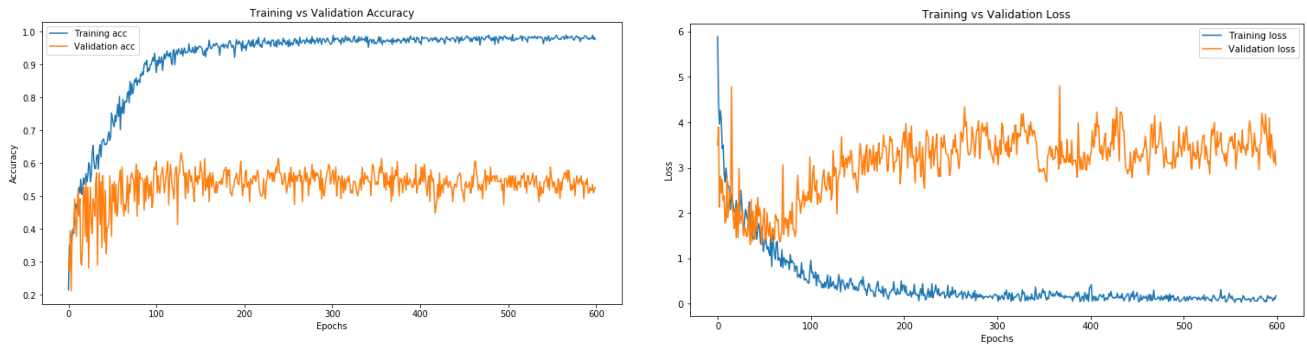

Figure 1. Accuracy and loss in training and validation after 600 epochs using 190 T1w sequences.

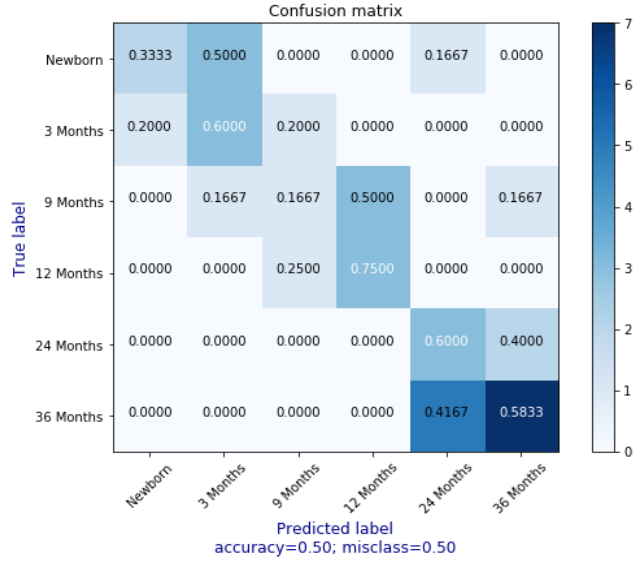

Figure 2. Relationship between neurodevelopmental age and predicted brain age using 190 T1w sequences.

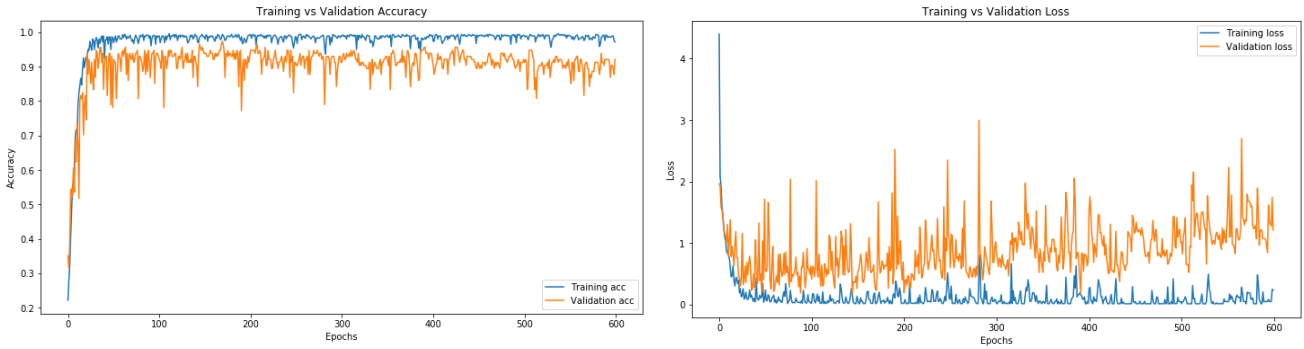

Figure 4. Accuracy and loss in training and validation after 600 epochs using 570 fusion MRI sequences after adding batch normalization.

| Class | Statistical Result of 3D CNN | | | Support |
|---|---|---|---|---|
| | *Precision* | *Recall* | *F1-score* | *N Scans* |
| newborn | 1.00 | 0.85 | 0.92 | 13 |
| 3 Months | 0.75 | 1.00 | 0.86 | 15 |
| 9 Months | 1.00 | 0.79 | 0.88 | 19 |
| 12 Months | 0.86 | 0.86 | 0.86 | 14 |
| 24 Months | 0.89 | 1.00 | 0.94 | 16 |
| 36 Months | 1.00 | 0.97 | 0.99 | 37 |

Table 1. 3D CNN statistical results after 600 epochs using 570 fusion MRI sequences after adding batch normalization.

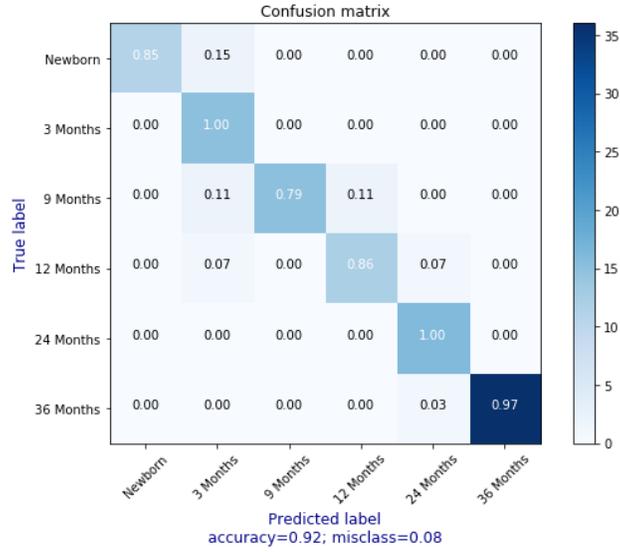

Figure 5. Relationship between neurodevelopmental age and predicted brain age in a 3D classification model after 600 epochs using 570 fusion MRI sequences after adding batch normalization.

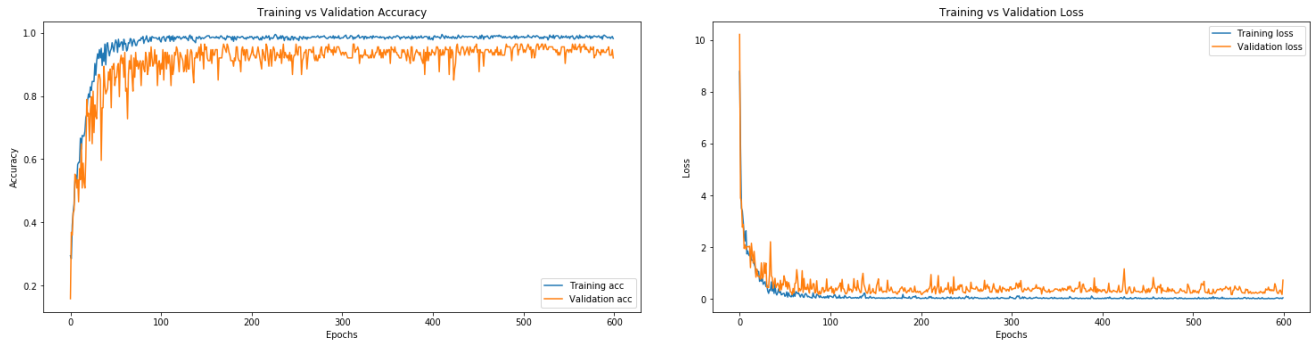

Figure 6. Accuracy and loss in training and validation after 600 epochs using 570 fusion MRI sequences after adding batch normalization and dropout.

| Class | Statistical Result of 3D CNN | | | Support |
|---|---|---|---|---|
| | *Precision* | *Recall* | *F1-score* | *N Scans* |
| newborn | 1.00 | 0.89 | 0.94 | 19 |
| 3 Months | 0.94 | 0.88 | 0.91 | 17 |
| 9 Months | 0.83 | 0.94 | 0.88 | 16 |
| 12 Months | 0.93 | 0.88 | 0.90 | 16 |
| 24 Months | 0.84 | 1.00 | 0.91 | 16 |
| 36 Months | 0.97 | 0.93 | 0.95 | 30 |

Table 2. 3D CNN statistical results after 600 epochs using 570 fusion MRI sequences after adding batch normalization and dropout.

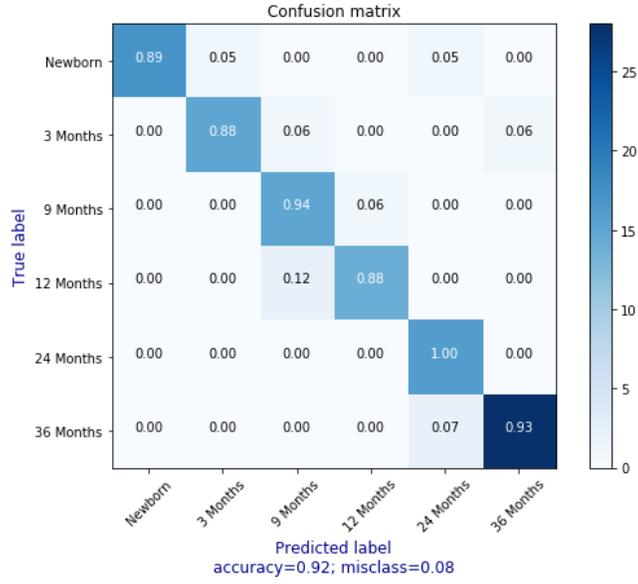

Figure 7. Relationship between neurodevelopmental age and predicted brain age in a 3D classification model after 600 epochs using 570 fusion MRI sequences after adding batch normalization and dropout.

| Models | Statistical Result using 570 T1w, T2w, and PDw sequences | | |
|---|---|---|---|
| | *Precision* | *Recall* | *Accuracy* |
| 3D CNN | 91.66% | 91.16% | 92% |
| 3D CNN+Batch Normalization | 91.88% | 92% | 92% |
| 3D CNN+ Batch Normalization+ dropout | 94.83% | 93.5% | 94% |

Table 3. Summary of statistical results with 570 fusion MRI sequences.